\documentclass[reprint,amsmath,amssymb,superscriptaddress,prl,floatfix,nobibnotes,showpacs]{revtex4-1}
\usepackage{graphicx}
\usepackage{bm}
\usepackage{float}

\begin{document}

\title{Distinct Charge Orders in the Planes and Chains of Ortho-III-Ordered YBa$_2$Cu$_3$O$_{6+\delta}$ Superconductors Identified by Resonant Elastic X-ray Scattering} 

\author{A. J. Achkar}
\affiliation{Department of Physics and Astronomy, University of Waterloo, Waterloo, Ontario, Canada N2L 3G1}
\author{R. Sutarto}
\affiliation{Department of Physics and Astronomy, University of British Columbia, Vancouver, British Columbia, Canada V6T 1Z4}
\affiliation{Canadian Light Source, University of Saskatchewan, Saskatoon, Saskatchewan, Canada S7N 0X4}
\author{X. Mao}
\affiliation{Department of Physics and Astronomy, University of Waterloo, Waterloo, Ontario, Canada N2L 3G1}
\author{F. He}
\affiliation{Canadian Light Source, University of Saskatchewan, Saskatoon, Saskatchewan, Canada S7N 0X4}
\author{A. Frano}
\affiliation{Max-Planck-Institut f\"{u}r Festk\"{o}rperforschung, Heisenbergstra\ss{}e 1, D-70569 Stuttgart, Germany}
\affiliation{Helmholtz-Zentrum Berlin f\"{u}r Materialien und Energie, Albert-Einstein-Stra\ss{}e 15, D-12489 Berlin, Germany}
\author{S. Blanco-Canosa}
\affiliation{Max-Planck-Institut f\"{u}r Festk\"{o}rperforschung, Heisenbergstra\ss{}e 1, D-70569 Stuttgart, Germany}
\author{M. Le Tacon}
\affiliation{Max-Planck-Institut f\"{u}r Festk\"{o}rperforschung, Heisenbergstra\ss{}e 1, D-70569 Stuttgart, Germany}
\author{G. Ghiringhelli}
\affiliation{CNR-SPIN, CNISM and Dipartimento di Fisica, Politecnico di Milano, Piazza Leonardo da Vinci 32, I-20133 Milano, Italy}
\author{L. Braicovich}
\affiliation{CNR-SPIN, CNISM and Dipartimento di Fisica, Politecnico di Milano, Piazza Leonardo da Vinci 32, I-20133 Milano, Italy}
\author{M. Minola}
\affiliation{CNR-SPIN, CNISM and Dipartimento di Fisica, Politecnico di Milano, Piazza Leonardo da Vinci 32, I-20133 Milano, Italy}
\author{M. Moretti Sala}
\affiliation{European Synchrotron Radiation Facility, BP 220, F-38043 Grenoble Cedex, France}
\author{C. Mazzoli}
\affiliation{CNR-SPIN, CNISM and Dipartimento di Fisica, Politecnico di Milano, Piazza Leonardo da Vinci 32, I-20133 Milano, Italy}
\author{Ruixing Liang}
\affiliation{Department of Physics and Astronomy, University of British Columbia, Vancouver, British Columbia, Canada V6T 1Z4}
\author{D. A. Bonn}
\affiliation{Department of Physics and Astronomy, University of British Columbia, Vancouver, British Columbia, Canada V6T 1Z4}
\author{W. N. Hardy}
\affiliation{Department of Physics and Astronomy, University of British Columbia, Vancouver, British Columbia, Canada V6T 1Z4}
\author{B. Keimer}
\affiliation{Max-Planck-Institut f\"{u}r Festk\"{o}rperforschung, Heisenbergstra\ss{}e 1, D-70569 Stuttgart, Germany}
\author{G. A. Sawatzky}
\affiliation{Department of Physics and Astronomy, University of British Columbia, Vancouver, British Columbia, Canada V6T 1Z4}
\author{D. G. Hawthorn}
\email[]{dhawthor@uwaterloo.ca.}
\affiliation{Department of Physics and Astronomy, University of Waterloo, Waterloo, Ontario, Canada N2L 3G1}

\begin{abstract}
Recently, charge density wave (CDW) order in the CuO$_2$ planes of underdoped YBa$_2$Cu$_3$O$_{6+\delta}$ was detected using resonant soft x-ray scattering. An important question remains: is the chain layer responsible for this charge ordering? Here, we explore the energy and polarization dependence of the resonant scattering intensity in a detwinned sample of YBa$_2$Cu$_3$O$_{6.75}$ with ortho-III oxygen ordering in the chain layer. We show that the ortho-III CDW order in the chains is distinct from the CDW order in the planes. The ortho-III structure gives rise to a commensurate superlattice reflection at $Q$=[0.33 0 $L$] whose energy and polarization dependence agrees with expectations for oxygen ordering and a spatial modulation of the Cu valence in the chains. Incommensurate peaks at [0.30 0 $L$] and [0 0.30 $L$] from the CDW order in the planes are shown to be distinct in $Q$ as well as their temperature, energy, and polarization dependence, and are thus unrelated to the structure of the chain layer. Moreover, the energy dependence of the CDW order in the planes is shown to result from a spatial modulation of energies of the Cu 2$p$ to 3$d_{x^2-y^2}$ transition, similar to stripe-ordered 214 cuprates.
\end{abstract}


\maketitle

Direct evidence for charge density wave (CDW) order in YBa$_2$Cu$_3$O$_{6+\delta}$ (YBCO) was recently observed in high magnetic field using nuclear magnetic resonance  \cite{Wu11} and in zero-field diffraction, first with resonant soft x-ray scattering (RSXS) \cite{Ghiringhelli12} and subsequently with hard x-ray scattering \cite{Chang12}.  Prior to these measurements, density wave order \cite{Kivelson03,Vojta09} had been observed in 214 cuprates [La$_{2-x-y}$(Ba,Sr)$_x$(Eu,Nd)$_y$CuO$_4$] \cite{Tranquada95} as well as  Ca$_{2-x}$Na$_x$CuO$_2$Cl$_2$ \cite{Hanaguri04} and Bi$_2$Sr$_2$CaCu$_2$O$_{8+\delta}$ \cite{Kohsaka08}. However, density wave order in YBCO---a material long considered a benchmark cuprate due to its low disorder and high $T_{c,\text{max}}\simeq 94.2$ K---had only been inferred indirectly, being offered as an explanation for Hall effect measurements \cite{LeBoeuf07} and the electron pockets observed in quantum oscillation experiments \cite{Doiron07,Millis07,Sebastian12}. The observation of density wave order in YBCO thus marks an important milestone in efforts to determine whether density wave order is generic to the cuprates while providing new opportunities to identify common features of CDW order in the cuprates.

RSXS is well suited to give direct insight into the nature of CDW order in YBCO. RSXS involves diffraction with the photon energy tuned through an x-ray absorption edge. This gives significant energy dependence to the atomic scattering form factor, $f(\omega)$, enhancing the scattering from weak ordering and providing sensitivity to the charge, spin and orbital occupation of specific elements. At the Cu $L$ absorption edge, the scattering is sensitive to modulations in the unoccupied Cu 3$d$ states that are central to the low energy physics of the cuprates \cite{Abbamonte05,Fink09,Fink11,Wilkins11,Achkar12}. The recent RSXS measurements of Ghiringhelli \textit{et al.} at the Cu $L$ absorption edge identified  superlattice peaks at $Q$ = [0.31 0 $L$] and [0 0.31 $L$] indicative of CDW order \cite{Ghiringhelli12}. They also showed that the intensity of the superlattice reflections peak at $\sim$$T_c$ and decrease in intensity for $T < T_c$, providing a clear link between the density wave order and superconductivity \cite{Ghiringhelli12}.  Importantly, based on the energy dependence of the scattering intensity and the presence of peaks at $H$=0.31 and $K$=0.31 in a detwinned sample, Ghiringhelli \textit{et al.} also demonstrate that the CDW superlattice peaks originate from modulations in the CuO$_2$ planes.

However, the possible role of the charge reservoir layer in stabilizing the CDW order is not yet clear.  In YBCO, the charge reservoir for the CuO$_2$ planes is composed of CuO chains. The Cu sites in the chains (Cu1) and planes (Cu2) have different orbital symmetries and contribute differently to x-ray absorption spectroscopy (XAS) and RSXS measurements \cite{Nucker95,Hawthorn11b}.   In addition to making the structure orthorhombic ($a \neq b$), the chain layer can be oxygen ordered into a variety of ``ortho'' ordered phases \cite{Beyers89,Zimmermann03}. For instance, the ortho-III phase corresponds to a repeated pattern of full--full--empty ordering of oxygen in the chains [see Fig.~1(a)] that produces a commensurate superlattice peak at [0.33 0 $L$], in close proximity to the [0.31 0 $L$] peaks observed in Ref.~\cite{Ghiringhelli12}.   In addition, the chains may also be susceptible to CDW order along the chains (producing incommensurate peaks at [0 $K$ $L$]) \cite{Edwards94,Grevin00,Derro02}. As such, the chains may  act to stabilize CDW order in YBCO akin to the low-temperature tetragonal (LTT) structural phase stabilizing spin and charge stripes in stripe ordered 214 cuprates \cite{Tranquada95,Fujita02b}.

\begin{figure}[b!]
\centering
\resizebox{\columnwidth}{!}{\includegraphics{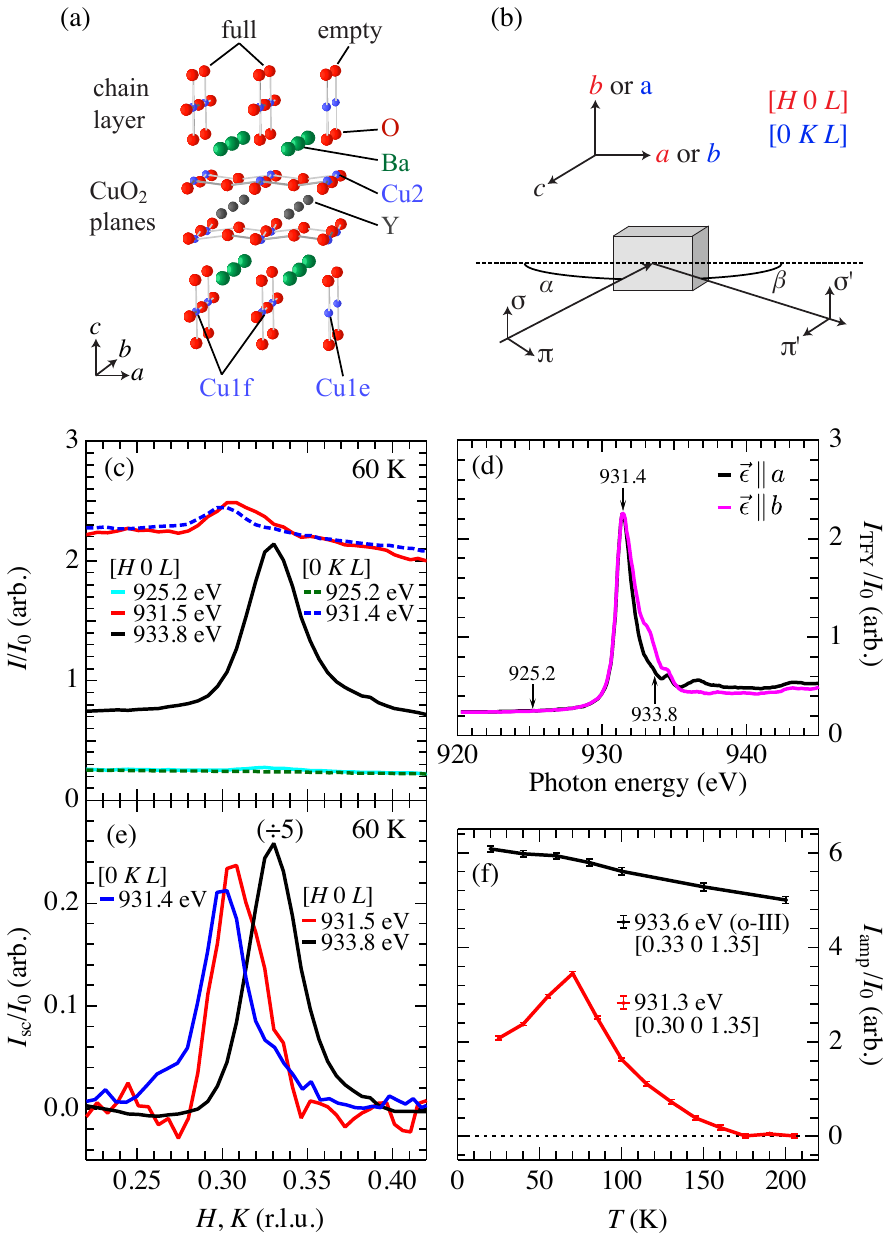}}
\caption{(color online) (a)  The crystal structure of ortho-III ordered YBCO.  (b) A schematic of the experiment geometry. (c) [$H$ 0 $L$] and [0 $K$ $L$] scans at $T$ = 60 K measured using sigma polarized light through the [0.30 0 1.4] and [0 0.30 1.4] superlattice peaks, which appear when the photon energy is tuned to the peak of the XAS ($\sim$931.4 eV).  The ortho-III oxygen ordering superlattice peak is seen at [0.33 0 $L$] and is most prominent around 933.8 eV.  (d) The x-ray absorption with polarization along the $a$ and $b$ axes measured using total fluorescence yield.  (e) The scattering intensity with fluorescence background subtracted.  (f) The temperature dependence of the amplitudes of the [0.30 0 1.35] and [0.33 0 1.35] peaks. r.l.u., reciprocal lattice units.}
\label{fig1}
\end{figure}

In this Letter, we present RSXS measurements of a high purity, ortho-III ordered single crystal of YBa$_2$Cu$_3$O$_{6.75}$ ($T_c$=75.2 K, $p$=0.133) \cite{Liang98,Liang06} that (a) confirm the in-plane origin of the incommensurate [0.30 0 $L$] CDW peak \footnote{The peaks in Ref.~\onlinecite{Ghiringhelli12}~are observed at $H$=0.31 and $K$=0.31 and not at $H$=0.30 and $K$=0.30 as in our measurements. At present, we cannot determine whether this difference is due to the different doping levels of the samples (a truly intrinsic effect) or some minor differences in the alignment of the crystals in the two studies.}, (b) clarify its relation to the oxygen ordering in the chain layer, and (c) demonstrate a link to the microscopic origin of stripes in 214 cuprates.  Analysis of the scattering intensities provides clear evidence that the [0.30 0 $L$] CDW peak has an energy, polarization, and temperature dependence that is distinct from the [0.33 0 $L$] oxygen ordering peak, indicating there is no clear relation between the chain layer and the [0.30 0 $L$] CDW order. Moreover, the [0.30 0 $L$] peak is shown to result from a spatial modulation of the energy of the Cu 2$p$ to 3$d_{x^2-y^2}$ transition, unlike the [0.33 0 $L$] oxygen ordering peak, which is described by a spatial modulation of the Cu valence.  The former is consistent with RSXS measurements in stripe ordered 214 cuprates \cite{Achkar12}, which is also described by the energy shift model, suggesting a common origin to the CDW order that is generic to the cuprates.  

Resonant scattering measurements were performed at the Canadian Light Source's Resonant Elastic and Inelastic X-ray Scattering (REIXS) beamline \cite{Hawthorn11a} using linearly polarized light in both $\sigma$ and $\pi$ scattering geometries, as depicted in Fig.~1(b). The sample orientation was confirmed by detection of [0 0 1],  [$\pm$1 0 2], and [0 $\pm$1 2] Bragg reflections at 2.05 keV.  XAS was measured using total fluorescence yield (TFY).

The measured intensity of $H$ and $K$ scans through the [0.30 0 1.4] and [0 0.30 1.4] peaks at 60 K is shown in Fig.~1(c) for the incident photon energies indicated in Fig.~1(d).  These superlattice reflections are observed above a large x-ray fluorescence background, similar to measurements from Ref.~\onlinecite{Ghiringhelli12}. In addition, there is also a peak at [0.33 0 $L$] that is evident at higher photon energy.  

The scattering intensity, $I_\text{sc}$, was determined by fitting the fluorescence background to a polynomial and subtracting it from the data [Fig.~1(e)].  This procedure was repeated as a function of photon energy and for both $\sigma$ and $\pi$ incident photon polarizations, as shown in Fig.~2.  In Fig.~2(a) and 2(b), two peaks at $H$=0.30 and $H$=0.33 are observed that resonate at different energies and have a different polarization dependence.  Due to the large width of both peaks, they overlap in $H$ forming one broad asymmetric peak, which is particularly evident around the peak in the x-ray absorption (931.4 eV).  In contrast, the peak at [0 0.30 1.4], shown in Fig.~2(c) and Fig.~2(d), resonates at 931.3 eV with only a small signature of the peak at $\sim$0.33, likely due to residual ($<$3\%) twinning of the sample.  From these scans at 60 K, the correlation lengths of the peaks are $\xi$($K$=0.30) $\simeq$ 42 \AA~, $\xi$($H$=0.30) $\simeq$ 40 \AA, and $\xi$($H$=0.33) $\simeq$ 37 \AA.  Consistent with previous work \cite{Ghiringhelli12,Chang12}, the amplitude of the [0.30 0 $L$] reflection is first distinguished from the fluorescence background at $\sim$160 K, peaks near $T_c$, and decreases for $T<T_c$, as shown in Fig.~1(f).  In contrast, the [0.33 0 $L$] peak amplitude exhibits a gradual $T$ dependence with no notable features at $T_c$ or 160 K.  

\begin{figure}
\centering
\resizebox{\columnwidth}{!}{\includegraphics{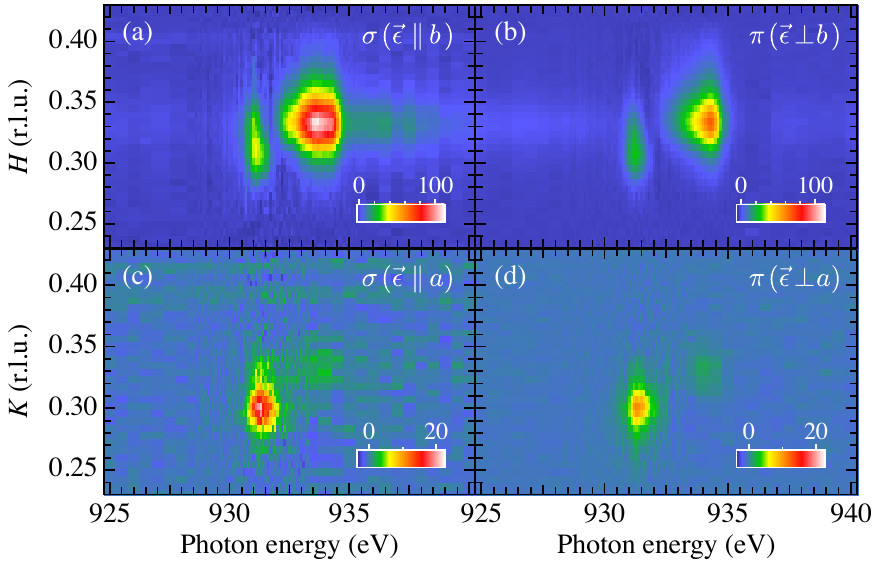}}
\caption{(color online) The [$H$ 0 $L$] [(a) and (b)] and [0 $K$ $L$] [(c) and (d)] normalized scattering intensity, $I_\text{sc}/I_0$, in arbitrary units.   The scattering intensity was measured with $\sigma$ [(a) and (c)] and $\pi$ [(b) and (d)] incident photon polarization at $T$ = 60 K. r.l.u., reciprocal lattice units.}
\label{fig2}
\end{figure}

\begin{figure}[b]
\centering
\resizebox{\columnwidth}{!}{\includegraphics{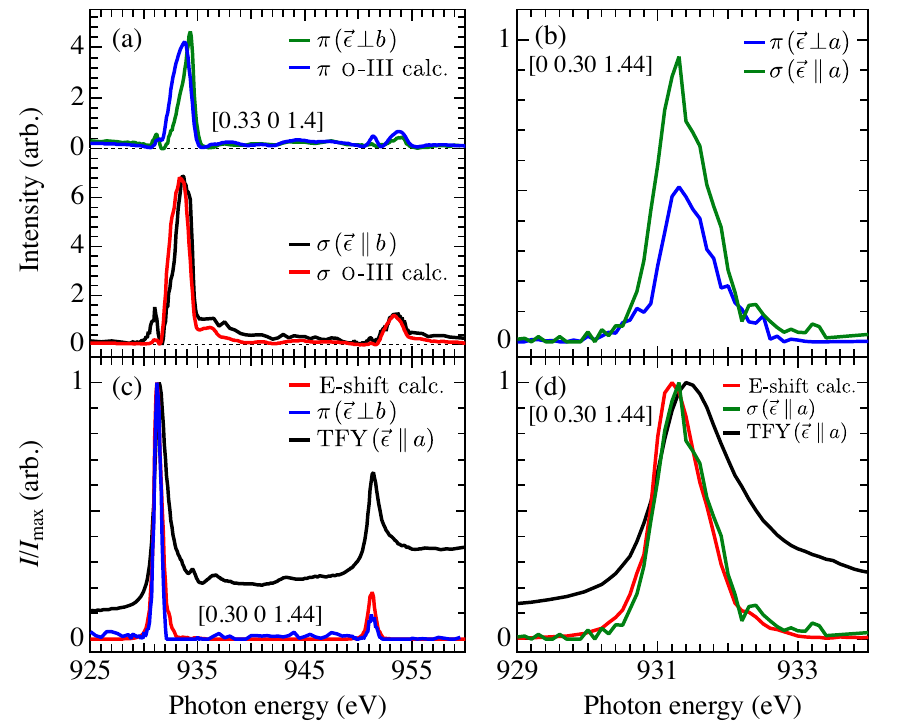}}
\caption{(color online) (a) The measured energy dependence of the [0.33 0 1.4] oxygen ordering peak with $\sigma$ and $\pi$ polarized incident light along with the calculated spectra for ortho-III oxygen ordering of the chain layer.  (b) The energy dependence of the [0 0.30 1.44] peak measured with $\sigma$ and $\pi$ polarized light.   (c) The energy dependence of the [0.30 0 1.44] peak (blue) with $\pi$ polarized light compared to the energy shift model calculation.  The energy shift calculation captures the correct peak position and energy width of the scattering intensity.  (d) The energy shift model calculation compared to the [0 0.30 1.44] peak with $\sigma$ polarized light.}
\label{fig3}
\end{figure}

Analysis of the energy and polarization dependence of the integrated scattering intensities (Fig.~3) demonstrates that the $H$=0.30 and $K$=0.30 peaks are due to modulations in the CuO$_2$ planes, whereas the $H$=0.33 peaks are due to ortho-III ordering in the chain layer. To model the scattering intensity of the $H$=0.33 peak, we followed the procedure in Ref.~\onlinecite{Hawthorn11b} which illustrated that the scattering intensity and polarization dependence of the oxygen order superstructure in ortho-II ordered YBCO (full--empty--full--empty chains) could be calculated by accounting for the impact of the oxygen dopants on the Cu1 $d$ states in the full and empty chains.   This was done by experimentally determining the energy dependence of the atomic scattering tensor, $F_i$, for Cu in full, $F_\text{Cu1f}(\omega)$, and empty, $F_\text{Cu1e}(\omega)$, chains using polarization dependent x-ray absorption measurements in YBCO prepared with either an entirely full (YBa$_2$Cu$_3$O$_7$) or an entirely empty (YBa$_2$Cu$_3$O$_6$) chain layer.  Here we use the same analysis for the $H$ = 0.33 peak with $F_\text{Cu1f}(\omega)$ and $F_\text{Cu1e}(\omega)$ from Ref.~\onlinecite{Hawthorn11b} and $I_\text{sc,o-III}(H\text{=0.33},\vec{\epsilon}) = |f_\text{Cu1f}(\omega,\vec{\epsilon}) +f_O-f_\text{Cu1e}(\omega,\vec{\epsilon})|^2$.  As shown in Fig.~3(a), this analysis reproduces the energy and polarization dependence of the $H$ = 0.33 peak, providing confirmation that this peak is dominated by the oxygen order in the chain layer.

In contrast, both the polarization and energy dependence of the $H$=0.30 and $K$=0.30 peaks are consistent with a spatial modulation of the Cu $3d_{x^2-y^2}$ states in the CuO$_2$ planes. First, one must note that the incident $\pi$ and $\sigma$ polarizations couple to different components of the scattering tensor.  For $\sigma$ polarization, the photon polarization is entirely along the $b$($a$) axis for the $H$($K$) = 0.30 peak and is therefore sensitive to the $bb$($aa$) components of the scattering tensor.  However, for $\pi$ polarized light, the polarization has components along both the $a$ and $c$ axes that depend on the scattering geometry.    For modulations of Cu $3d_{x^2-y^2}$ states, $f_{aa,\text{Cu}2} \simeq f_{bb,\text{Cu}2} >> f_{cc,\text{Cu}2}$ and  $I_\text{sc}(\pi \pi^\prime)/I_\text{sc}(\sigma\sigma^{\prime}) = [\sin(\alpha)\sin(\beta)\Delta f_{aa}]^2$, where $\alpha$ and $\beta$ are the angles of the incident and scattered light relative to the sample surface [see Fig.~1(b)] \footnote{We assume only $\sigma \sigma^\prime$ or $\pi \pi^\prime$ scattering is allowed (no magnetic or anisotropic tensor susceptibility scattering.)}.  For the values of $\alpha$ and $\beta$ in our measurement, one would expect the ratio of $I_\text{sc}(\pi \pi^{\prime})/I_\text{sc}(\sigma\sigma^{\prime}) = 0.46$ for a modulation of Cu $3d_{x^2-y^2}$ states.  As shown in Fig.~3(b), the $K$ = 0.30 peak is in good agreement with this ratio.

A final intriguing aspect of the energy dependence of the scattering intensity is that the line shape can be described by a simple phenomenological model for the scattering intensity based on a spatial modulation of the {\it energy} of the Cu 2$p$ to $3d_{x^2-y^2}$ transition.  The energy of this transition is determined by the energy of the $3d_{x^2-y^2}$ states, as well as the core hole energy and the interaction energy of the core hole with the $d$ electrons, all of which may be spatially modulated.   This energy shift model was recently shown to account for the energy dependence of the scattering intensity of the [1/4 0 $L$] charge stripe ordering peak in La$_{1.475}$Nd$_{0.4}$Sr$_{0.125}$CuO$_4$, unlike models based on lattice displacements or charge density modulations \cite{Achkar12}.  Although in YBCO we do not know the structure factor that accounts for the [0.30 0 $L$] and [0 0.30 $L$] peaks, we can naively invoke the same energy shift model and assume that $I_\text{sc}[0.30~0~L](\omega) \propto I_\text{sc}[0~0.30~L](\omega) \propto |f_{\text{Cu2a}}(\hbar \omega + \Delta E) - f_{\text{Cu2b}}(\hbar \omega - \Delta E)|^2$, where Cu2a and Cu2b represent two sites in the CuO$_2$ planes with $f(\omega)$ that is identical apart from a small energy shift $\pm \Delta E$ at each site.  Following previous work, $f_{\text{Cu}2}(\omega)$ can be determined from the experimentally measured x-ray absorption spectra \cite{Abbamonte05,Fink09,Achkar12}. In this case, the XAS with polarization oriented along the $a$-axis of the sample is used since it is dominated by the Cu 3$d_{x^2-y^2}$ states of the CuO$_2$ planes with minimal chain contribution.  As shown in Fig.~3(c) and Fig.~3(d), the energy shift model is in excellent agreement with the experiment, capturing the correct energy dependence and peak position, which peaks $\sim$$0.1$ eV below the $L_3$ peak of the x-ray absorption. Note, for this calculation $\Delta E$=0.1 eV was used (see~\footnote{As discussed in Ref.~\onlinecite{Achkar12}, the line shape is insensitive to the magnitude of $\Delta E$ provided $\Delta E <0.2$ eV. As such, $\Delta E$ should not be considered a fitting parameter.}).

Although the energy shifts, and thus the scattering, may ultimately be caused by a modulation in Cu valence, the microscopic origin of the energy shifts is currently unclear.   An important implication of the energy shift model is that the resonant scattering provides only indirect evidence for charge density (valence) modulations---the success of the energy shift model allows one to infer there is a charge density modulation since this must occur if the electronic structure is spatially modulated \cite{Achkar12}.  In contrast, the energy dependence of the ortho-III oxygen order peak ($H$ = 0.33) is described ``directly" in terms of a large change in valence between Cu in the full and empty chains.   However, since we cannot presently estimate the magnitude of the charge density modulation from the energy shifts, it is conceivable that a modulation of charge is not the central feature of the newfound density wave order in YBCO (and also stripes in 214 cuprates).  In such a case, the energy shifts may in fact be a signature of a novel electronic state, such as a valence bond solid \cite{Achkar12}.   Alternately, the energy shifts may result from weak-coupling, Fermi surface reconstruction descriptions of density wave order in the cuprates. Regardless of the origin, the success of the energy shift model may imply that the temperature dependence of the peak amplitudes results from a temperature dependent energy shift that peaks at $T_c$, providing an apparent link between the energy shifts and superconductivity.

Moreover, the applicability of the energy shift model to the resonant scattering intensity of charge stripe order in 214 cuprates and YBCO indicates that the CDW order likely shares a common origin in the two material systems.  This commonality stands in contrast to important differences between the density wave order in YBCO and stripe ordered 214 cuprates.  In 214 cuprates, the charge order is stabilized by the LTT structural phase \cite{Tranquada95,Fujita02}, has an incommensurability that plateaus at high doping at the commensurate value of $2\delta$ = 0.25 \cite{Yamada98,Vojta09} and is understood to be unidirectional in nature (i.e. stripes). In YBCO, while there is no LTT phase, one might expect that the orthorhombic structure of YBCO would preferentially stabilize stripe order propagating along the $a$ or $b$ axes, perhaps with a period locked to the oxygen ordering in the chain layer of YBCO.  However, no clear link between structure and the $H$ and $K$ = 0.30 peaks is observed in our measurements.  Rather, the incommensurate value of the $2\delta$ = 0.30 peaks relative to the commensurate oxygen ordering peak at $H$ = 0.33, the similar magnitude of the scattering intensity of the $H$ and $K$ peaks and the presence of the $H$ = 0.30 peaks in samples with weak oxygen order (only very short range ortho-V order) \cite{Ghiringhelli12}, indicate that the structural distortions are not an essential ingredient for CDW order in YBCO.  Additionally, the existence of peaks along both $H$ and $K$ would seem to favor 2D checkerboard order.  However, if the connection to the lattice is indeed weak, domains of unidirectional stripes oriented along both $a$ and $b$ may describe the CDW order in YBCO.

Finally, in addition to structural distortions, which may provide pinning centres commensurate with the lattice, disorder can also provide random pinning centres for density wave order and has been shown to enhance spin density wave and CDW order in 214 cuprates \cite{Hirota01,Fujita08,Suchaneck10}.  Moreover, 214 cuprates are intrinsically disordered owing to the chemical cation substitution (ex. Sr for La) near the CuO$_2$ planes used to dope away from half filling.  This makes it difficult to disentangle the role of disorder from the intrinsic physics of 214 cuprates. In contrast, the presence of CDW order in high-purity, oxygen ordered YBCO provides a strong indication that density wave order is in fact an intrinsic feature of underdoped cuprates.

\begin{acknowledgements}
We thank T. Senthil, A. Burkov, S. Wilkins and K. Shen for discussions. This work was supported by the Canada Foundation for Innovation, the Canadian Institute for Advanced Research, and the Natural Sciences and Engineering Research Council of Canada.  The research described in this Letter was performed at the Canadian Light Source, which is supported by NSERC, NRC, CIHR, and the University of Saskatchewan.
\end{acknowledgements}


\begin{thebibliography}{36}%
\makeatletter
\providecommand \@ifxundefined [1]{%
 \@ifx{#1\undefined}
}%
\providecommand \@ifnum [1]{%
 \ifnum #1\expandafter \@firstoftwo
 \else \expandafter \@secondoftwo
 \fi
}%
\providecommand \@ifx [1]{%
 \ifx #1\expandafter \@firstoftwo
 \else \expandafter \@secondoftwo
 \fi
}%
\providecommand \natexlab [1]{#1}%
\providecommand \enquote  [1]{``#1''}%
\providecommand \bibnamefont  [1]{#1}%
\providecommand \bibfnamefont [1]{#1}%
\providecommand \citenamefont [1]{#1}%
\providecommand \href@noop [0]{\@secondoftwo}%
\providecommand \href [0]{\begingroup \@sanitize@url \@href}%
\providecommand \@href[1]{\@@startlink{#1}\@@href}%
\providecommand \@@href[1]{\endgroup#1\@@endlink}%
\providecommand \@sanitize@url [0]{\catcode `\\12\catcode `\$12\catcode
  `\&12\catcode `\#12\catcode `\^12\catcode `\_12\catcode `\%12\relax}%
\providecommand \@@startlink[1]{}%
\providecommand \@@endlink[0]{}%
\providecommand \url  [0]{\begingroup\@sanitize@url \@url }%
\providecommand \@url [1]{\endgroup\@href {#1}{\urlprefix }}%
\providecommand \urlprefix  [0]{URL }%
\providecommand \Eprint [0]{\href }%
\@ifxundefined \urlstyle {%
  \providecommand \doi  [0]{\begingroup \@sanitize@url \@doi}%
  \providecommand \@doi [1]{\endgroup \@@startlink {\doibase
  #1}doi:\discretionary {}{}{}#1\@@endlink }%
}{%
  \providecommand \doi  [0]{doi:\discretionary{}{}{}\begingroup
  \urlstyle{rm}\Url }%
}%
\providecommand \doibase [0]{http://dx.doi.org/}%
\providecommand \Doi [0]{\begingroup \@sanitize@url \@Doi }%
\providecommand \@Doi  [1]{\endgroup\@@startlink{\doibase#1}\@@Doi}%
\providecommand \@@Doi [1]{#1\@@endlink}%
\providecommand \selectlanguage [0]{\@gobble}%
\providecommand \bibinfo  [0]{\@secondoftwo}%
\providecommand \bibfield  [0]{\@secondoftwo}%
\providecommand \translation [1]{[#1]}%
\providecommand \BibitemOpen [0]{}%
\providecommand \bibitemStop [0]{}%
\providecommand \bibitemNoStop [0]{.\EOS\space}%
\providecommand \EOS [0]{\spacefactor3000\relax}%
\providecommand \BibitemShut  [1]{\csname bibitem#1\endcsname}%
\bibitem [{\citenamefont {Wu}\ \emph {et~al.}(2011)\citenamefont {Wu},
  \citenamefont {Mayaffre}, \citenamefont {Kr{\"a}mer}, \citenamefont
  {Horvati{\'c}}, \citenamefont {Berthier}, \citenamefont {Hardy},
  \citenamefont {Liang}, \citenamefont {Bonn},\ and\ \citenamefont
  {Julien}}]{Wu11}%
  \BibitemOpen
  \bibfield  {author} {\bibinfo {author} {\bibfnamefont {T.}~\bibnamefont
  {Wu}}, \bibinfo {author} {\bibfnamefont {H.}~\bibnamefont {Mayaffre}},
  \bibinfo {author} {\bibfnamefont {S.}~\bibnamefont {Kr{\"a}mer}}, \bibinfo
  {author} {\bibfnamefont {M.}~\bibnamefont {Horvati{\'c}}}, \bibinfo {author}
  {\bibfnamefont {C.}~\bibnamefont {Berthier}}, \bibinfo {author}
  {\bibfnamefont {W.~N.}\ \bibnamefont {Hardy}}, \bibinfo {author}
  {\bibfnamefont {R.}~\bibnamefont {Liang}}, \bibinfo {author} {\bibfnamefont
  {D.~A.}\ \bibnamefont {Bonn}}, \ and\ \bibinfo {author} {\bibfnamefont
  {M.-H.}\ \bibnamefont {Julien}},\ }\href@noop {} {\bibfield  {journal}
  {\bibinfo  {journal} {Nature},\ }\textbf {\bibinfo {volume} {477}},\ \bibinfo
  {pages} {191} (\bibinfo {year} {2011})}\BibitemShut {NoStop}%
\bibitem [{\citenamefont {Ghiringhelli}\ \emph {et~al.}(2012)\citenamefont
  {Ghiringhelli}, \citenamefont {Le~Tacon}, \citenamefont {Minola},
  \citenamefont {Blanco-Canosa}, \citenamefont {Mazzoli}, \citenamefont
  {Brookes}, \citenamefont {De~Luca}, \citenamefont {Frano}, \citenamefont
  {Hawthorn}, \citenamefont {He}, \citenamefont {Loew}, \citenamefont {Sala},
  \citenamefont {Peets}, \citenamefont {Salluzzo}, \citenamefont {Schierle},
  \citenamefont {Sutarto}, \citenamefont {Sawatzky}, \citenamefont {Weschke},
  \citenamefont {Keimer},\ and\ \citenamefont {Braicovich}}]{Ghiringhelli12}%
  \BibitemOpen
  \bibfield  {author} {\bibinfo {author} {\bibfnamefont {G.}~\bibnamefont
  {Ghiringhelli}}, \bibinfo {author} {\bibfnamefont {M.}~\bibnamefont
  {Le~Tacon}}, \bibinfo {author} {\bibfnamefont {M.}~\bibnamefont {Minola}},
  \bibinfo {author} {\bibfnamefont {S.}~\bibnamefont {Blanco-Canosa}}, \bibinfo
  {author} {\bibfnamefont {C.}~\bibnamefont {Mazzoli}}, \bibinfo {author}
  {\bibfnamefont {N.~B.}\ \bibnamefont {Brookes}}, \bibinfo {author}
  {\bibfnamefont {G.~M.}\ \bibnamefont {De~Luca}}, \bibinfo {author}
  {\bibfnamefont {A.}~\bibnamefont {Frano}}, \bibinfo {author} {\bibfnamefont
  {D.~G.}\ \bibnamefont {Hawthorn}}, \bibinfo {author} {\bibfnamefont
  {F.}~\bibnamefont {He}}, \bibinfo {author} {\bibfnamefont {T.}~\bibnamefont
  {Loew}}, \bibinfo {author} {\bibfnamefont {M.~M.}\ \bibnamefont {Sala}},
  \bibinfo {author} {\bibfnamefont {D.~C.}\ \bibnamefont {Peets}}, \bibinfo
  {author} {\bibfnamefont {M.}~\bibnamefont {Salluzzo}}, \bibinfo {author}
  {\bibfnamefont {E.}~\bibnamefont {Schierle}}, \bibinfo {author}
  {\bibfnamefont {R.}~\bibnamefont {Sutarto}}, \bibinfo {author} {\bibfnamefont
  {G.~A.}\ \bibnamefont {Sawatzky}}, \bibinfo {author} {\bibfnamefont
  {E.}~\bibnamefont {Weschke}}, \bibinfo {author} {\bibfnamefont
  {B.}~\bibnamefont {Keimer}}, \ and\ \bibinfo {author} {\bibfnamefont
  {L.}~\bibnamefont {Braicovich}},\ }\href@noop {} {\bibfield  {journal}
  {\bibinfo  {journal} {Science},\ }\textbf {\bibinfo {volume} {337}},\
  \bibinfo {pages} {821} (\bibinfo {year} {2012})}\BibitemShut {NoStop}%
\bibitem [{\citenamefont {Chang}\ \emph {et~al.}()\citenamefont {Chang},
  \citenamefont {Blackburn}, \citenamefont {Holmes}, \citenamefont
  {Christensen}, \citenamefont {Larsen}, \citenamefont {Mesot}, \citenamefont
  {Liang}, \citenamefont {Bonn}, \citenamefont {Hardy}, \citenamefont
  {Watenphul}, \citenamefont {Von~Zimmermann}, \citenamefont {Forgan},\ and\
  \citenamefont {Hayden}}]{Chang12}%
  \BibitemOpen
  \bibfield  {author} {\bibinfo {author} {\bibfnamefont {J.}~\bibnamefont
  {Chang}}, \bibinfo {author} {\bibfnamefont {E.}~\bibnamefont {Blackburn}},
  \bibinfo {author} {\bibfnamefont {A.~T.}\ \bibnamefont {Holmes}}, \bibinfo
  {author} {\bibfnamefont {N.~B.}\ \bibnamefont {Christensen}}, \bibinfo
  {author} {\bibfnamefont {J.}~\bibnamefont {Larsen}}, \bibinfo {author}
  {\bibfnamefont {J.}~\bibnamefont {Mesot}}, \bibinfo {author} {\bibfnamefont
  {R.}~\bibnamefont {Liang}}, \bibinfo {author} {\bibfnamefont {D.~A.}\
  \bibnamefont {Bonn}}, \bibinfo {author} {\bibfnamefont {W.~N.}\ \bibnamefont
  {Hardy}}, \bibinfo {author} {\bibfnamefont {A.}~\bibnamefont {Watenphul}},
  \bibinfo {author} {\bibfnamefont {M.}~\bibnamefont {Von~Zimmermann}},
  \bibinfo {author} {\bibfnamefont {E.~M.}\ \bibnamefont {Forgan}}, \ and\
  \bibinfo {author} {\bibfnamefont {S.~M.}\ \bibnamefont {Hayden}},\
  }\href@noop {} {}\bibinfo {note} {arXiv:1206.4333}\BibitemShut {NoStop}%
\bibitem [{\citenamefont {Kivelson}\ \emph {et~al.}(2003)\citenamefont
  {Kivelson}, \citenamefont {Bindloss}, \citenamefont {Fradkin}, \citenamefont
  {Oganesyan}, \citenamefont {Tranquada}, \citenamefont {Kapitulnik},\ and\
  \citenamefont {Howald}}]{Kivelson03}%
  \BibitemOpen
  \bibfield  {author} {\bibinfo {author} {\bibfnamefont {S.~A.}\ \bibnamefont
  {Kivelson}}, \bibinfo {author} {\bibfnamefont {I.~P.}\ \bibnamefont
  {Bindloss}}, \bibinfo {author} {\bibfnamefont {E.}~\bibnamefont {Fradkin}},
  \bibinfo {author} {\bibfnamefont {V.}~\bibnamefont {Oganesyan}}, \bibinfo
  {author} {\bibfnamefont {J.~M.}\ \bibnamefont {Tranquada}}, \bibinfo {author}
  {\bibfnamefont {A.}~\bibnamefont {Kapitulnik}}, \ and\ \bibinfo {author}
  {\bibfnamefont {C.}~\bibnamefont {Howald}},\ }\href@noop {} {\bibfield
  {journal} {\bibinfo  {journal} {Rev. Mod. Phys.},\ }\textbf {\bibinfo
  {volume} {75}},\ \bibinfo {pages} {1201} (\bibinfo {year}
  {2003})}\BibitemShut {NoStop}%
\bibitem [{\citenamefont {Vojta}(2009)}]{Vojta09}%
  \BibitemOpen
  \bibfield  {author} {\bibinfo {author} {\bibfnamefont {M.}~\bibnamefont
  {Vojta}},\ }\href@noop {} {\bibfield  {journal} {\bibinfo  {journal} {Adv.
  Phys.},\ }\textbf {\bibinfo {volume} {58}},\ \bibinfo {pages} {699} (\bibinfo
  {year} {2009})}\BibitemShut {NoStop}%
\bibitem [{\citenamefont {Tranquada}\ \emph {et~al.}(1995)\citenamefont
  {Tranquada}, \citenamefont {Sternlieb}, \citenamefont {Axe}, \citenamefont
  {Nakamura},\ and\ \citenamefont {Uchida}}]{Tranquada95}%
  \BibitemOpen
  \bibfield  {author} {\bibinfo {author} {\bibfnamefont {J.~M.}\ \bibnamefont
  {Tranquada}}, \bibinfo {author} {\bibfnamefont {B.~J.}\ \bibnamefont
  {Sternlieb}}, \bibinfo {author} {\bibfnamefont {J.~D.}\ \bibnamefont {Axe}},
  \bibinfo {author} {\bibfnamefont {Y.}~\bibnamefont {Nakamura}}, \ and\
  \bibinfo {author} {\bibfnamefont {S.}~\bibnamefont {Uchida}},\ }\href@noop {}
  {\bibfield  {journal} {\bibinfo  {journal} {Nature},\ }\textbf {\bibinfo
  {volume} {375}},\ \bibinfo {pages} {561} (\bibinfo {year}
  {1995})}\BibitemShut {NoStop}%
\bibitem [{\citenamefont {Hanaguri}\ \emph {et~al.}(2004)\citenamefont
  {Hanaguri}, \citenamefont {Lupien}, \citenamefont {Kohsaka}, \citenamefont
  {Lee}, \citenamefont {Azuma}, \citenamefont {Takano}, \citenamefont
  {Takagi},\ and\ \citenamefont {Davis}}]{Hanaguri04}%
  \BibitemOpen
  \bibfield  {author} {\bibinfo {author} {\bibfnamefont {T.}~\bibnamefont
  {Hanaguri}}, \bibinfo {author} {\bibfnamefont {C.}~\bibnamefont {Lupien}},
  \bibinfo {author} {\bibfnamefont {Y.}~\bibnamefont {Kohsaka}}, \bibinfo
  {author} {\bibfnamefont {D.-H.}\ \bibnamefont {Lee}}, \bibinfo {author}
  {\bibfnamefont {M.}~\bibnamefont {Azuma}}, \bibinfo {author} {\bibfnamefont
  {M.}~\bibnamefont {Takano}}, \bibinfo {author} {\bibfnamefont
  {H.}~\bibnamefont {Takagi}}, \ and\ \bibinfo {author} {\bibfnamefont
  {J.}~\bibnamefont {Davis}},\ }\href@noop {} {\bibfield  {journal} {\bibinfo
  {journal} {Nature},\ }\textbf {\bibinfo {volume} {430}},\ \bibinfo {pages}
  {1001} (\bibinfo {year} {2004})}\BibitemShut {NoStop}%
\bibitem [{\citenamefont {Kohsaka}\ \emph {et~al.}(2008)\citenamefont
  {Kohsaka}, \citenamefont {Taylor}, \citenamefont {Wahl}, \citenamefont
  {Schmidt}, \citenamefont {Lee}, \citenamefont {Fujita}, \citenamefont
  {Alldredge}, \citenamefont {McElroy}, \citenamefont {Lee}, \citenamefont
  {Eisaki}, \citenamefont {Uchida}, \citenamefont {Lee},\ and\ \citenamefont
  {Davis}}]{Kohsaka08}%
  \BibitemOpen
  \bibfield  {author} {\bibinfo {author} {\bibfnamefont {Y.}~\bibnamefont
  {Kohsaka}}, \bibinfo {author} {\bibfnamefont {C.}~\bibnamefont {Taylor}},
  \bibinfo {author} {\bibfnamefont {P.}~\bibnamefont {Wahl}}, \bibinfo {author}
  {\bibfnamefont {A.}~\bibnamefont {Schmidt}}, \bibinfo {author} {\bibfnamefont
  {J.}~\bibnamefont {Lee}}, \bibinfo {author} {\bibfnamefont {K.}~\bibnamefont
  {Fujita}}, \bibinfo {author} {\bibfnamefont {J.~W.}\ \bibnamefont
  {Alldredge}}, \bibinfo {author} {\bibfnamefont {K.}~\bibnamefont {McElroy}},
  \bibinfo {author} {\bibfnamefont {J.}~\bibnamefont {Lee}}, \bibinfo {author}
  {\bibfnamefont {H.}~\bibnamefont {Eisaki}}, \bibinfo {author} {\bibfnamefont
  {S.}~\bibnamefont {Uchida}}, \bibinfo {author} {\bibfnamefont {D.-H.}\
  \bibnamefont {Lee}}, \ and\ \bibinfo {author} {\bibfnamefont {J.~C.}\
  \bibnamefont {Davis}},\ }\href@noop {} {\bibfield  {journal} {\bibinfo
  {journal} {Nature},\ }\textbf {\bibinfo {volume} {454}},\ \bibinfo {pages}
  {1072} (\bibinfo {year} {2008})}\BibitemShut {NoStop}%
\bibitem [{\citenamefont {LeBoeuf}\ \emph {et~al.}(2007)\citenamefont
  {LeBoeuf}, \citenamefont {Doiron-Leyraud}, \citenamefont {Levallois},
  \citenamefont {Daou}, \citenamefont {Bonnemaison}, \citenamefont {Hussey},
  \citenamefont {Balicas}, \citenamefont {Ramshaw}, \citenamefont {Liang},
  \citenamefont {Bonn}, \citenamefont {Hardy}, \citenamefont {Adachi},
  \citenamefont {Proust},\ and\ \citenamefont {Taillefer}}]{LeBoeuf07}%
  \BibitemOpen
  \bibfield  {author} {\bibinfo {author} {\bibfnamefont {D.}~\bibnamefont
  {LeBoeuf}}, \bibinfo {author} {\bibfnamefont {N.}~\bibnamefont
  {Doiron-Leyraud}}, \bibinfo {author} {\bibfnamefont {J.}~\bibnamefont
  {Levallois}}, \bibinfo {author} {\bibfnamefont {R.}~\bibnamefont {Daou}},
  \bibinfo {author} {\bibfnamefont {J.-B.}\ \bibnamefont {Bonnemaison}},
  \bibinfo {author} {\bibfnamefont {N.~E.}\ \bibnamefont {Hussey}}, \bibinfo
  {author} {\bibfnamefont {L.}~\bibnamefont {Balicas}}, \bibinfo {author}
  {\bibfnamefont {B.~J.}\ \bibnamefont {Ramshaw}}, \bibinfo {author}
  {\bibfnamefont {R.}~\bibnamefont {Liang}}, \bibinfo {author} {\bibfnamefont
  {D.~A.}\ \bibnamefont {Bonn}}, \bibinfo {author} {\bibfnamefont {W.~N.}\
  \bibnamefont {Hardy}}, \bibinfo {author} {\bibfnamefont {S.}~\bibnamefont
  {Adachi}}, \bibinfo {author} {\bibfnamefont {C.}~\bibnamefont {Proust}}, \
  and\ \bibinfo {author} {\bibfnamefont {L.}~\bibnamefont {Taillefer}},\
  }\href@noop {} {\bibfield  {journal} {\bibinfo  {journal} {Nature},\ }\textbf
  {\bibinfo {volume} {450}},\ \bibinfo {pages} {533} (\bibinfo {year}
  {2007})}\BibitemShut {NoStop}%
\bibitem [{\citenamefont {Doiron-Leyraud}\ \emph {et~al.}(2007)\citenamefont
  {Doiron-Leyraud}, \citenamefont {Proust}, \citenamefont {LeBoeuf},
  \citenamefont {Levallois}, \citenamefont {Bonnemaison}, \citenamefont
  {Liang}, \citenamefont {Bonn}, \citenamefont {Hardy},\ and\ \citenamefont
  {Taillefer}}]{Doiron07}%
  \BibitemOpen
  \bibfield  {author} {\bibinfo {author} {\bibfnamefont {N.}~\bibnamefont
  {Doiron-Leyraud}}, \bibinfo {author} {\bibfnamefont {C.}~\bibnamefont
  {Proust}}, \bibinfo {author} {\bibfnamefont {D.}~\bibnamefont {LeBoeuf}},
  \bibinfo {author} {\bibfnamefont {J.}~\bibnamefont {Levallois}}, \bibinfo
  {author} {\bibfnamefont {J.-B.}\ \bibnamefont {Bonnemaison}}, \bibinfo
  {author} {\bibfnamefont {R.}~\bibnamefont {Liang}}, \bibinfo {author}
  {\bibfnamefont {D.~A.}\ \bibnamefont {Bonn}}, \bibinfo {author}
  {\bibfnamefont {W.~N.}\ \bibnamefont {Hardy}}, \ and\ \bibinfo {author}
  {\bibfnamefont {L.}~\bibnamefont {Taillefer}},\ }\href@noop {} {\bibfield
  {journal} {\bibinfo  {journal} {Nature},\ }\textbf {\bibinfo {volume}
  {447}},\ \bibinfo {pages} {563} (\bibinfo {year} {2007})}\BibitemShut
  {NoStop}%
\bibitem [{\citenamefont {Millis}\ and\ \citenamefont
  {Norman}(2007)}]{Millis07}%
  \BibitemOpen
  \bibfield  {author} {\bibinfo {author} {\bibfnamefont {A.~J.}\ \bibnamefont
  {Millis}}\ and\ \bibinfo {author} {\bibfnamefont {M.~R.}\ \bibnamefont
  {Norman}},\ }\href {http://link.aps.org/doi/10.1103/PhysRevB.76.220503}
  {\bibfield  {journal} {\bibinfo  {journal} {Phys. Rev. B},\ }\textbf
  {\bibinfo {volume} {76}},\ \bibinfo {pages} {220503} (\bibinfo {year}
  {2007})}\BibitemShut {NoStop}%
\bibitem [{\citenamefont {Sebastian}\ \emph {et~al.}(2012)\citenamefont
  {Sebastian}, \citenamefont {Harrison},\ and\ \citenamefont
  {Lonzarich}}]{Sebastian12}%
  \BibitemOpen
  \bibfield  {author} {\bibinfo {author} {\bibfnamefont {S.~E.}\ \bibnamefont
  {Sebastian}}, \bibinfo {author} {\bibfnamefont {N.}~\bibnamefont {Harrison}},
  \ and\ \bibinfo {author} {\bibfnamefont {G.~G.}\ \bibnamefont {Lonzarich}},\
  }\href {http://dx.doi.org/10.1088/0034-4885/75/10/102501} {\bibfield
  {journal} {\bibinfo  {journal} {Reports on Progress in Physics},\ }\textbf
  {\bibinfo {volume} {75}},\ \bibinfo {pages} {102501} (\bibinfo {year}
  {2012})}\BibitemShut {NoStop}%
\bibitem [{\citenamefont {Abbamonte}\ \emph {et~al.}(2005)\citenamefont
  {Abbamonte}, \citenamefont {Rusydi}, \citenamefont {Smadici}, \citenamefont
  {Gu}, \citenamefont {Sawatzky},\ and\ \citenamefont {Feng}}]{Abbamonte05}%
  \BibitemOpen
  \bibfield  {author} {\bibinfo {author} {\bibfnamefont {P.}~\bibnamefont
  {Abbamonte}}, \bibinfo {author} {\bibfnamefont {A.}~\bibnamefont {Rusydi}},
  \bibinfo {author} {\bibfnamefont {S.}~\bibnamefont {Smadici}}, \bibinfo
  {author} {\bibfnamefont {G.~D.}\ \bibnamefont {Gu}}, \bibinfo {author}
  {\bibfnamefont {G.~A.}\ \bibnamefont {Sawatzky}}, \ and\ \bibinfo {author}
  {\bibfnamefont {D.~L.}\ \bibnamefont {Feng}},\ }\href@noop {} {\bibfield
  {journal} {\bibinfo  {journal} {Nat. Phys.},\ }\textbf {\bibinfo {volume}
  {1}},\ \bibinfo {pages} {155} (\bibinfo {year} {2005})}\BibitemShut {NoStop}%
\bibitem [{\citenamefont {Fink}\ \emph {et~al.}(2009)\citenamefont {Fink},
  \citenamefont {Schierle}, \citenamefont {Weschke}, \citenamefont {Geck},
  \citenamefont {Hawthorn}, \citenamefont {Soltwisch}, \citenamefont {Wadati},
  \citenamefont {Wu}, \citenamefont {Durr}, \citenamefont {Wizent},
  \citenamefont {Buchner},\ and\ \citenamefont {Sawatzky}}]{Fink09}%
  \BibitemOpen
  \bibfield  {author} {\bibinfo {author} {\bibfnamefont {J.}~\bibnamefont
  {Fink}}, \bibinfo {author} {\bibfnamefont {E.}~\bibnamefont {Schierle}},
  \bibinfo {author} {\bibfnamefont {E.}~\bibnamefont {Weschke}}, \bibinfo
  {author} {\bibfnamefont {J.}~\bibnamefont {Geck}}, \bibinfo {author}
  {\bibfnamefont {D.}~\bibnamefont {Hawthorn}}, \bibinfo {author}
  {\bibfnamefont {V.}~\bibnamefont {Soltwisch}}, \bibinfo {author}
  {\bibfnamefont {H.}~\bibnamefont {Wadati}}, \bibinfo {author} {\bibfnamefont
  {H.-H.}\ \bibnamefont {Wu}}, \bibinfo {author} {\bibfnamefont {H.~A.}\
  \bibnamefont {Durr}}, \bibinfo {author} {\bibfnamefont {N.}~\bibnamefont
  {Wizent}}, \bibinfo {author} {\bibfnamefont {B.}~\bibnamefont {Buchner}}, \
  and\ \bibinfo {author} {\bibfnamefont {G.~A.}\ \bibnamefont {Sawatzky}},\
  }\href@noop {} {\bibfield  {journal} {\bibinfo  {journal} {Phys. Rev. B},\
  }\textbf {\bibinfo {volume} {79}},\ \bibinfo {eid} {100502} (\bibinfo {year}
  {2009})}\BibitemShut {NoStop}%
\bibitem [{\citenamefont {Fink}\ \emph {et~al.}(2011)\citenamefont {Fink},
  \citenamefont {Soltwisch}, \citenamefont {Geck}, \citenamefont {Schierle},
  \citenamefont {Weschke},\ and\ \citenamefont {B{\"u}chner}}]{Fink11}%
  \BibitemOpen
  \bibfield  {author} {\bibinfo {author} {\bibfnamefont {J.}~\bibnamefont
  {Fink}}, \bibinfo {author} {\bibfnamefont {V.}~\bibnamefont {Soltwisch}},
  \bibinfo {author} {\bibfnamefont {J.}~\bibnamefont {Geck}}, \bibinfo {author}
  {\bibfnamefont {E.}~\bibnamefont {Schierle}}, \bibinfo {author}
  {\bibfnamefont {E.}~\bibnamefont {Weschke}}, \ and\ \bibinfo {author}
  {\bibfnamefont {B.}~\bibnamefont {B{\"u}chner}},\ }\href@noop {} {\bibfield
  {journal} {\bibinfo  {journal} {Phys. Rev. B},\ }\textbf {\bibinfo {volume}
  {83}},\ \bibinfo {pages} {092503} (\bibinfo {year} {2011})}\BibitemShut
  {NoStop}%
\bibitem [{\citenamefont {Wilkins}\ \emph {et~al.}(2011)\citenamefont
  {Wilkins}, \citenamefont {Dean}, \citenamefont {Fink}, \citenamefont
  {H\"ucker}, \citenamefont {Geck}, \citenamefont {Soltwisch}, \citenamefont
  {Schierle}, \citenamefont {Weschke}, \citenamefont {Gu}, \citenamefont
  {Uchida}, \citenamefont {Ichikawa}, \citenamefont {Tranquada},\ and\
  \citenamefont {Hill}}]{Wilkins11}%
  \BibitemOpen
  \bibfield  {author} {\bibinfo {author} {\bibfnamefont {S.~B.}\ \bibnamefont
  {Wilkins}}, \bibinfo {author} {\bibfnamefont {M.~P.~M.}\ \bibnamefont
  {Dean}}, \bibinfo {author} {\bibfnamefont {J.}~\bibnamefont {Fink}}, \bibinfo
  {author} {\bibfnamefont {M.}~\bibnamefont {H\"ucker}}, \bibinfo {author}
  {\bibfnamefont {J.}~\bibnamefont {Geck}}, \bibinfo {author} {\bibfnamefont
  {V.}~\bibnamefont {Soltwisch}}, \bibinfo {author} {\bibfnamefont
  {E.}~\bibnamefont {Schierle}}, \bibinfo {author} {\bibfnamefont
  {E.}~\bibnamefont {Weschke}}, \bibinfo {author} {\bibfnamefont
  {G.}~\bibnamefont {Gu}}, \bibinfo {author} {\bibfnamefont {S.}~\bibnamefont
  {Uchida}}, \bibinfo {author} {\bibfnamefont {N.}~\bibnamefont {Ichikawa}},
  \bibinfo {author} {\bibfnamefont {J.~M.}\ \bibnamefont {Tranquada}}, \ and\
  \bibinfo {author} {\bibfnamefont {J.~P.}\ \bibnamefont {Hill}},\ }\href@noop
  {} {\bibfield  {journal} {\bibinfo  {journal} {Phys. Rev. B},\ }\textbf
  {\bibinfo {volume} {84}},\ \bibinfo {pages} {195101} (\bibinfo {year}
  {2011})}\BibitemShut {NoStop}%
\bibitem [{\citenamefont {Achkar}\ \emph {et~al.}()\citenamefont {Achkar},
  \citenamefont {He}, \citenamefont {Sutarto}, \citenamefont {Geck},
  \citenamefont {Zhang}, \citenamefont {Kim},\ and\ \citenamefont
  {Hawthorn}}]{Achkar12}%
  \BibitemOpen
  \bibfield  {author} {\bibinfo {author} {\bibfnamefont {A.~J.}\ \bibnamefont
  {Achkar}}, \bibinfo {author} {\bibfnamefont {F.}~\bibnamefont {He}}, \bibinfo
  {author} {\bibfnamefont {R.}~\bibnamefont {Sutarto}}, \bibinfo {author}
  {\bibfnamefont {J.}~\bibnamefont {Geck}}, \bibinfo {author} {\bibfnamefont
  {H.}~\bibnamefont {Zhang}}, \bibinfo {author} {\bibfnamefont {Y.-J.}\
  \bibnamefont {Kim}}, \ and\ \bibinfo {author} {\bibfnamefont {D.~G.}\
  \bibnamefont {Hawthorn}},\ }\href@noop {} {}\bibinfo {note}
  {arXiv:1203.2669}\BibitemShut {NoStop}%
\bibitem [{\citenamefont {N\"ucker}\ \emph {et~al.}(1995)\citenamefont
  {N\"ucker}, \citenamefont {Pellegrin}, \citenamefont {Schweiss},
  \citenamefont {Fink}, \citenamefont {Molodtsov}, \citenamefont {Simmons},
  \citenamefont {Kaindl}, \citenamefont {Frentrup}, \citenamefont {Erb},\ and\
  \citenamefont {M\"uller-Vogt}}]{Nucker95}%
  \BibitemOpen
  \bibfield  {author} {\bibinfo {author} {\bibfnamefont {N.}~\bibnamefont
  {N\"ucker}}, \bibinfo {author} {\bibfnamefont {E.}~\bibnamefont {Pellegrin}},
  \bibinfo {author} {\bibfnamefont {P.}~\bibnamefont {Schweiss}}, \bibinfo
  {author} {\bibfnamefont {J.}~\bibnamefont {Fink}}, \bibinfo {author}
  {\bibfnamefont {S.~L.}\ \bibnamefont {Molodtsov}}, \bibinfo {author}
  {\bibfnamefont {C.~T.}\ \bibnamefont {Simmons}}, \bibinfo {author}
  {\bibfnamefont {G.}~\bibnamefont {Kaindl}}, \bibinfo {author} {\bibfnamefont
  {W.}~\bibnamefont {Frentrup}}, \bibinfo {author} {\bibfnamefont
  {A.}~\bibnamefont {Erb}}, \ and\ \bibinfo {author} {\bibfnamefont
  {G.}~\bibnamefont {M\"uller-Vogt}},\ }\href@noop {} {\bibfield  {journal}
  {\bibinfo  {journal} {Phys. Rev. B},\ }\textbf {\bibinfo {volume} {51}},\
  \bibinfo {pages} {8529} (\bibinfo {year} {1995})}\BibitemShut {NoStop}%
\bibitem [{\citenamefont {Hawthorn}\ \emph
  {et~al.}(2011){\natexlab{a}}\citenamefont {Hawthorn}, \citenamefont {Shen},
  \citenamefont {Geck}, \citenamefont {Peets}, \citenamefont {Wadati},
  \citenamefont {Okamoto}, \citenamefont {Huang}, \citenamefont {Huang},
  \citenamefont {Lin}, \citenamefont {Denlinger}, \citenamefont {Liang},
  \citenamefont {Bonn}, \citenamefont {Hardy},\ and\ \citenamefont
  {Sawatzky}}]{Hawthorn11b}%
  \BibitemOpen
  \bibfield  {author} {\bibinfo {author} {\bibfnamefont {D.~G.}\ \bibnamefont
  {Hawthorn}}, \bibinfo {author} {\bibfnamefont {K.~M.}\ \bibnamefont {Shen}},
  \bibinfo {author} {\bibfnamefont {J.}~\bibnamefont {Geck}}, \bibinfo {author}
  {\bibfnamefont {D.~C.}\ \bibnamefont {Peets}}, \bibinfo {author}
  {\bibfnamefont {H.}~\bibnamefont {Wadati}}, \bibinfo {author} {\bibfnamefont
  {J.}~\bibnamefont {Okamoto}}, \bibinfo {author} {\bibfnamefont {S.-W.}\
  \bibnamefont {Huang}}, \bibinfo {author} {\bibfnamefont {D.~J.}\ \bibnamefont
  {Huang}}, \bibinfo {author} {\bibfnamefont {H.-J.}\ \bibnamefont {Lin}},
  \bibinfo {author} {\bibfnamefont {J.~D.}\ \bibnamefont {Denlinger}}, \bibinfo
  {author} {\bibfnamefont {R.}~\bibnamefont {Liang}}, \bibinfo {author}
  {\bibfnamefont {D.~A.}\ \bibnamefont {Bonn}}, \bibinfo {author}
  {\bibfnamefont {W.~N.}\ \bibnamefont {Hardy}}, \ and\ \bibinfo {author}
  {\bibfnamefont {G.~A.}\ \bibnamefont {Sawatzky}},\ }\href@noop {} {\bibfield
  {journal} {\bibinfo  {journal} {Phys. Rev. B},\ }\textbf {\bibinfo {volume}
  {84}},\ \bibinfo {pages} {075125} (\bibinfo {year}
  {2011}{\natexlab{a}})}\BibitemShut {NoStop}%
\bibitem [{\citenamefont {Beyers}\ \emph {et~al.}(1989)\citenamefont {Beyers},
  \citenamefont {Ahn}, \citenamefont {Gorman}, \citenamefont {Lee},
  \citenamefont {Parkin}, \citenamefont {Ramirez}, \citenamefont {Roche},
  \citenamefont {Vazquez}, \citenamefont {G{\"u}r},\ and\ \citenamefont
  {Huggins}}]{Beyers89}%
  \BibitemOpen
  \bibfield  {author} {\bibinfo {author} {\bibfnamefont {R.}~\bibnamefont
  {Beyers}}, \bibinfo {author} {\bibfnamefont {B.~T.}\ \bibnamefont {Ahn}},
  \bibinfo {author} {\bibfnamefont {G.}~\bibnamefont {Gorman}}, \bibinfo
  {author} {\bibfnamefont {V.~Y.}\ \bibnamefont {Lee}}, \bibinfo {author}
  {\bibfnamefont {S.~S.~P.}\ \bibnamefont {Parkin}}, \bibinfo {author}
  {\bibfnamefont {M.~L.}\ \bibnamefont {Ramirez}}, \bibinfo {author}
  {\bibfnamefont {K.~P.}\ \bibnamefont {Roche}}, \bibinfo {author}
  {\bibfnamefont {J.~E.}\ \bibnamefont {Vazquez}}, \bibinfo {author}
  {\bibfnamefont {T.~M.}\ \bibnamefont {G{\"u}r}}, \ and\ \bibinfo {author}
  {\bibfnamefont {R.~A.}\ \bibnamefont {Huggins}},\ }\href@noop {} {\bibfield
  {journal} {\bibinfo  {journal} {Nature},\ }\textbf {\bibinfo {volume}
  {340}},\ \bibinfo {pages} {619} (\bibinfo {year} {1989})}\BibitemShut
  {NoStop}%
\bibitem [{\citenamefont {v.~Zimmermann}\ \emph {et~al.}(2003)\citenamefont
  {v.~Zimmermann}, \citenamefont {Schneider}, \citenamefont {Frello},
  \citenamefont {Andersen}, \citenamefont {Madsen}, \citenamefont {ll},
  \citenamefont {Poulsen}, \citenamefont {Liang}, \citenamefont {Dosanjh},\
  and\ \citenamefont {Hardy}}]{Zimmermann03}%
  \BibitemOpen
  \bibfield  {author} {\bibinfo {author} {\bibfnamefont {M.}~\bibnamefont
  {v.~Zimmermann}}, \bibinfo {author} {\bibfnamefont {J.~R.}\ \bibnamefont
  {Schneider}}, \bibinfo {author} {\bibfnamefont {T.}~\bibnamefont {Frello}},
  \bibinfo {author} {\bibfnamefont {N.~H.}\ \bibnamefont {Andersen}}, \bibinfo
  {author} {\bibfnamefont {J.}~\bibnamefont {Madsen}}, \bibinfo {author}
  {\bibfnamefont {M.~K.}\ \bibnamefont {ll}}, \bibinfo {author} {\bibfnamefont
  {H.~F.}\ \bibnamefont {Poulsen}}, \bibinfo {author} {\bibfnamefont
  {R.}~\bibnamefont {Liang}}, \bibinfo {author} {\bibfnamefont
  {P.}~\bibnamefont {Dosanjh}}, \ and\ \bibinfo {author} {\bibfnamefont
  {W.~N.}\ \bibnamefont {Hardy}},\ }\href@noop {} {\bibfield  {journal}
  {\bibinfo  {journal} {Phys. Rev. B},\ }\textbf {\bibinfo {volume} {68}},\
  \bibinfo {pages} {104515} (\bibinfo {year} {2003})}\BibitemShut {NoStop}%
\bibitem [{\citenamefont {Edwards}\ \emph {et~al.}(1994)\citenamefont
  {Edwards}, \citenamefont {Barr}, \citenamefont {Markert},\ and\ \citenamefont
  {de~Lozanne}}]{Edwards94}%
  \BibitemOpen
  \bibfield  {author} {\bibinfo {author} {\bibfnamefont {H.~L.}\ \bibnamefont
  {Edwards}}, \bibinfo {author} {\bibfnamefont {A.~L.}\ \bibnamefont {Barr}},
  \bibinfo {author} {\bibfnamefont {J.~T.}\ \bibnamefont {Markert}}, \ and\
  \bibinfo {author} {\bibfnamefont {A.~L.}\ \bibnamefont {de~Lozanne}},\
  }\href@noop {} {\bibfield  {journal} {\bibinfo  {journal} {Phys. Rev.
  Lett.},\ }\textbf {\bibinfo {volume} {73}},\ \bibinfo {pages} {1154}
  (\bibinfo {year} {1994})}\BibitemShut {NoStop}%
\bibitem [{\citenamefont {Gr{\'e}vin}\ \emph {et~al.}(2000)\citenamefont
  {Gr{\'e}vin}, \citenamefont {Berthier},\ and\ \citenamefont
  {Collin}}]{Grevin00}%
  \BibitemOpen
  \bibfield  {author} {\bibinfo {author} {\bibfnamefont {B.}~\bibnamefont
  {Gr{\'e}vin}}, \bibinfo {author} {\bibfnamefont {Y.}~\bibnamefont
  {Berthier}}, \ and\ \bibinfo {author} {\bibfnamefont {G.}~\bibnamefont
  {Collin}},\ }\href@noop {} {\bibfield  {journal} {\bibinfo  {journal} {Phys.
  Rev. Lett.},\ }\textbf {\bibinfo {volume} {85}},\ \bibinfo {pages} {1310}
  (\bibinfo {year} {2000})}\BibitemShut {NoStop}%
\bibitem [{\citenamefont {Derro}\ \emph {et~al.}(2002)\citenamefont {Derro},
  \citenamefont {Hudson}, \citenamefont {Lang}, \citenamefont {Pan},
  \citenamefont {Davis}, \citenamefont {Markert},\ and\ \citenamefont
  {de~Lozanne}}]{Derro02}%
  \BibitemOpen
  \bibfield  {author} {\bibinfo {author} {\bibfnamefont {D.~J.}\ \bibnamefont
  {Derro}}, \bibinfo {author} {\bibfnamefont {E.~W.}\ \bibnamefont {Hudson}},
  \bibinfo {author} {\bibfnamefont {K.~M.}\ \bibnamefont {Lang}}, \bibinfo
  {author} {\bibfnamefont {S.~H.}\ \bibnamefont {Pan}}, \bibinfo {author}
  {\bibfnamefont {J.~C.}\ \bibnamefont {Davis}}, \bibinfo {author}
  {\bibfnamefont {J.~T.}\ \bibnamefont {Markert}}, \ and\ \bibinfo {author}
  {\bibfnamefont {A.~L.}\ \bibnamefont {de~Lozanne}},\ }\href@noop {}
  {\bibfield  {journal} {\bibinfo  {journal} {Phys. Rev. Lett.},\ }\textbf
  {\bibinfo {volume} {88}},\ \bibinfo {pages} {097002} (\bibinfo {year}
  {2002})}\BibitemShut {NoStop}%
\bibitem [{\citenamefont {Fujita}\ \emph
  {et~al.}(2002){\natexlab{a}}\citenamefont {Fujita}, \citenamefont {Goka},
  \citenamefont {Yamada},\ and\ \citenamefont {Matsuda}}]{Fujita02b}%
  \BibitemOpen
  \bibfield  {author} {\bibinfo {author} {\bibfnamefont {M.}~\bibnamefont
  {Fujita}}, \bibinfo {author} {\bibfnamefont {H.}~\bibnamefont {Goka}},
  \bibinfo {author} {\bibfnamefont {K.}~\bibnamefont {Yamada}}, \ and\ \bibinfo
  {author} {\bibfnamefont {M.}~\bibnamefont {Matsuda}},\ }\href@noop {}
  {\bibfield  {journal} {\bibinfo  {journal} {Phys. Rev. Lett.},\ }\textbf
  {\bibinfo {volume} {88}},\ \bibinfo {pages} {167008} (\bibinfo {year}
  {2002}{\natexlab{a}})}\BibitemShut {NoStop}%
\bibitem [{\citenamefont {Liang}\ \emph {et~al.}(1998)\citenamefont {Liang},
  \citenamefont {Bonn},\ and\ \citenamefont {Hardy}}]{Liang98}%
  \BibitemOpen
  \bibfield  {author} {\bibinfo {author} {\bibfnamefont {R.}~\bibnamefont
  {Liang}}, \bibinfo {author} {\bibfnamefont {D.~A.}\ \bibnamefont {Bonn}}, \
  and\ \bibinfo {author} {\bibfnamefont {W.~N.}\ \bibnamefont {Hardy}},\
  }\href@noop {} {\bibfield  {journal} {\bibinfo  {journal} {Physica C},\
  }\textbf {\bibinfo {volume} {304}},\ \bibinfo {pages} {105} (\bibinfo {year}
  {1998})}\BibitemShut {NoStop}%
\bibitem [{\citenamefont {Liang}\ \emph {et~al.}(2006)\citenamefont {Liang},
  \citenamefont {Bonn},\ and\ \citenamefont {Hardy}}]{Liang06}%
  \BibitemOpen
  \bibfield  {author} {\bibinfo {author} {\bibfnamefont {R.}~\bibnamefont
  {Liang}}, \bibinfo {author} {\bibfnamefont {D.~A.}\ \bibnamefont {Bonn}}, \
  and\ \bibinfo {author} {\bibfnamefont {W.~N.}\ \bibnamefont {Hardy}},\
  }\href@noop {} {\bibfield  {journal} {\bibinfo  {journal} {Phys. Rev. B},\
  }\textbf {\bibinfo {volume} {73}},\ \bibinfo {pages} {180505} (\bibinfo
  {year} {2006})}\BibitemShut {NoStop}%
\bibitem [{Note1()}]{Note1}%
  \BibitemOpen
  \bibinfo {note} {The peaks in Ref.~\protect \rev@citealp {Ghiringhelli12}~are
  observed at $H$=0.31 and $K$=0.31 and not at $H$=0.30 and $K$=0.30 as in our
  measurements. At present, we cannot determine whether this difference is due
  to the different doping levels of the samples (a truly intrinsic effect) or
  some minor differences in the alignment of the crystals in the two
  studies.}\BibitemShut {Stop}%
\bibitem [{\citenamefont {Hawthorn}\ \emph
  {et~al.}(2011){\natexlab{b}}\citenamefont {Hawthorn}, \citenamefont {He},
  \citenamefont {Venema}, \citenamefont {Davis}, \citenamefont {Achkar},
  \citenamefont {Zhang}, \citenamefont {Sutarto}, \citenamefont {Wadati},
  \citenamefont {Radi}, \citenamefont {Wilson}, \citenamefont {Wright},
  \citenamefont {Shen}, \citenamefont {Geck}, \citenamefont {Zhang},
  \citenamefont {Nov{\'a}k},\ and\ \citenamefont {Sawatzky}}]{Hawthorn11a}%
  \BibitemOpen
  \bibfield  {author} {\bibinfo {author} {\bibfnamefont {D.~G.}\ \bibnamefont
  {Hawthorn}}, \bibinfo {author} {\bibfnamefont {F.}~\bibnamefont {He}},
  \bibinfo {author} {\bibfnamefont {L.}~\bibnamefont {Venema}}, \bibinfo
  {author} {\bibfnamefont {H.}~\bibnamefont {Davis}}, \bibinfo {author}
  {\bibfnamefont {A.~J.}\ \bibnamefont {Achkar}}, \bibinfo {author}
  {\bibfnamefont {J.}~\bibnamefont {Zhang}}, \bibinfo {author} {\bibfnamefont
  {R.}~\bibnamefont {Sutarto}}, \bibinfo {author} {\bibfnamefont
  {H.}~\bibnamefont {Wadati}}, \bibinfo {author} {\bibfnamefont
  {A.}~\bibnamefont {Radi}}, \bibinfo {author} {\bibfnamefont {T.}~\bibnamefont
  {Wilson}}, \bibinfo {author} {\bibfnamefont {G.}~\bibnamefont {Wright}},
  \bibinfo {author} {\bibfnamefont {K.~M.}\ \bibnamefont {Shen}}, \bibinfo
  {author} {\bibfnamefont {J.}~\bibnamefont {Geck}}, \bibinfo {author}
  {\bibfnamefont {H.}~\bibnamefont {Zhang}}, \bibinfo {author} {\bibfnamefont
  {V.}~\bibnamefont {Nov{\'a}k}}, \ and\ \bibinfo {author} {\bibfnamefont
  {G.~A.}\ \bibnamefont {Sawatzky}},\ }\href@noop {} {\bibfield  {journal}
  {\bibinfo  {journal} {Rev. Sci. Instrum.},\ }\textbf {\bibinfo {volume}
  {82}},\ \bibinfo {pages} {073104} (\bibinfo {year}
  {2011}{\natexlab{b}})}\BibitemShut {NoStop}%
\bibitem [{Note2()}]{Note2}%
  \BibitemOpen
  \bibinfo {note} {We assume only $\sigma \sigma ^\prime $ or $\pi \pi ^\prime
  $ scattering is allowed (no magnetic or anisotropic tensor susceptibility scattering.)}\BibitemShut
  {NoStop}%
\bibitem [{Note3()}]{Note3}%
  \BibitemOpen
  \bibinfo {note} {As discussed in Ref.~\protect \rev@citealp {Achkar12}, the
  line shape is insensitive to the magnitude of $\Delta E$ provided $\Delta E
  <0.2$ eV. As such, $\Delta E$ should not be considered a fitting
  parameter.}\BibitemShut {Stop}%
\bibitem [{\citenamefont {Fujita}\ \emph
  {et~al.}(2002){\natexlab{b}}\citenamefont {Fujita}, \citenamefont {Yamada},
  \citenamefont {Hiraka}, \citenamefont {Gehring}, \citenamefont {Lee},
  \citenamefont {Wakimoto},\ and\ \citenamefont {Shirane}}]{Fujita02}%
  \BibitemOpen
  \bibfield  {author} {\bibinfo {author} {\bibfnamefont {M.}~\bibnamefont
  {Fujita}}, \bibinfo {author} {\bibfnamefont {K.}~\bibnamefont {Yamada}},
  \bibinfo {author} {\bibfnamefont {H.}~\bibnamefont {Hiraka}}, \bibinfo
  {author} {\bibfnamefont {P.~M.}\ \bibnamefont {Gehring}}, \bibinfo {author}
  {\bibfnamefont {S.~H.}\ \bibnamefont {Lee}}, \bibinfo {author} {\bibfnamefont
  {S.}~\bibnamefont {Wakimoto}}, \ and\ \bibinfo {author} {\bibfnamefont
  {G.}~\bibnamefont {Shirane}},\ }\href@noop {} {\bibfield  {journal} {\bibinfo
   {journal} {Phys. Rev. B},\ }\textbf {\bibinfo {volume} {65}},\ \bibinfo
  {pages} {064505} (\bibinfo {year} {2002}{\natexlab{b}})}\BibitemShut
  {NoStop}%
\bibitem [{\citenamefont {Yamada}\ \emph {et~al.}(1998)\citenamefont {Yamada},
  \citenamefont {Lee}, \citenamefont {Kurahashi}, \citenamefont {Wada},
  \citenamefont {Wakimoto}, \citenamefont {Ueki}, \citenamefont {Kimura},
  \citenamefont {Endoh}, \citenamefont {Hosoya}, \citenamefont {Shirane},
  \citenamefont {Birgeneau}, \citenamefont {Greven}, \citenamefont {Kastner},\
  and\ \citenamefont {Kim}}]{Yamada98}%
  \BibitemOpen
  \bibfield  {author} {\bibinfo {author} {\bibfnamefont {K.}~\bibnamefont
  {Yamada}}, \bibinfo {author} {\bibfnamefont {C.~H.}\ \bibnamefont {Lee}},
  \bibinfo {author} {\bibfnamefont {K.}~\bibnamefont {Kurahashi}}, \bibinfo
  {author} {\bibfnamefont {J.}~\bibnamefont {Wada}}, \bibinfo {author}
  {\bibfnamefont {S.}~\bibnamefont {Wakimoto}}, \bibinfo {author}
  {\bibfnamefont {S.}~\bibnamefont {Ueki}}, \bibinfo {author} {\bibfnamefont
  {H.}~\bibnamefont {Kimura}}, \bibinfo {author} {\bibfnamefont
  {Y.}~\bibnamefont {Endoh}}, \bibinfo {author} {\bibfnamefont
  {S.}~\bibnamefont {Hosoya}}, \bibinfo {author} {\bibfnamefont
  {G.}~\bibnamefont {Shirane}}, \bibinfo {author} {\bibfnamefont {R.~J.}\
  \bibnamefont {Birgeneau}}, \bibinfo {author} {\bibfnamefont {M.}~\bibnamefont
  {Greven}}, \bibinfo {author} {\bibfnamefont {M.~A.}\ \bibnamefont {Kastner}},
  \ and\ \bibinfo {author} {\bibfnamefont {Y.~J.}\ \bibnamefont {Kim}},\ }\href
  {http://link.aps.org/%doi/10.1103/PhysRevB.57.6165} {\bibfield  {journal}
  {\bibinfo  {journal} {Phys. Rev. B},\ }\textbf {\bibinfo {volume} {57}},\
  \bibinfo {pages} {6165} (\bibinfo {year} {1998})}\BibitemShut {NoStop}%
\bibitem [{\citenamefont {Hirota}(2001)}]{Hirota01}%
  \BibitemOpen
  \bibfield  {author} {\bibinfo {author} {\bibfnamefont {K.}~\bibnamefont
  {Hirota}},\ }\href
  {http://www.sciencedirect.com/science/article/pii/S0921453401001952}
  {\bibfield  {journal} {\bibinfo  {journal} {Physica C},\ }\textbf {\bibinfo
  {volume} {357 - 360}},\ \bibinfo {pages} {61 } (\bibinfo {year}
  {2001})}\BibitemShut {NoStop}%
\bibitem [{\citenamefont {Fujita}\ \emph {et~al.}(2008)\citenamefont {Fujita},
  \citenamefont {Enoki},\ and\ \citenamefont {Yamada}}]{Fujita08}%
  \BibitemOpen
  \bibfield  {author} {\bibinfo {author} {\bibfnamefont {M.}~\bibnamefont
  {Fujita}}, \bibinfo {author} {\bibfnamefont {M.}~\bibnamefont {Enoki}}, \
  and\ \bibinfo {author} {\bibfnamefont {K.}~\bibnamefont {Yamada}},\ }\href
  {http://www.sciencedirect.com/science/article/pii/S0022369708002667}
  {\bibfield  {journal} {\bibinfo  {journal} {J. Phys. Chem. Sol.},\ }\textbf
  {\bibinfo {volume} {69}},\ \bibinfo {pages} {3167 } (\bibinfo {year}
  {2008})}\BibitemShut {NoStop}%
\bibitem [{\citenamefont {Suchaneck}\ \emph {et~al.}(2010)\citenamefont
  {Suchaneck}, \citenamefont {Hinkov}, \citenamefont {Haug}, \citenamefont
  {Schulz}, \citenamefont {Bernhard}, \citenamefont {Ivanov}, \citenamefont
  {Hradil}, \citenamefont {Lin}, \citenamefont {Bourges}, \citenamefont
  {Keimer},\ and\ \citenamefont {Sidis}}]{Suchaneck10}%
  \BibitemOpen
  \bibfield  {author} {\bibinfo {author} {\bibfnamefont {A.}~\bibnamefont
  {Suchaneck}}, \bibinfo {author} {\bibfnamefont {V.}~\bibnamefont {Hinkov}},
  \bibinfo {author} {\bibfnamefont {D.}~\bibnamefont {Haug}}, \bibinfo {author}
  {\bibfnamefont {L.}~\bibnamefont {Schulz}}, \bibinfo {author} {\bibfnamefont
  {C.}~\bibnamefont {Bernhard}}, \bibinfo {author} {\bibfnamefont
  {A.}~\bibnamefont {Ivanov}}, \bibinfo {author} {\bibfnamefont
  {K.}~\bibnamefont {Hradil}}, \bibinfo {author} {\bibfnamefont {C.~T.}\
  \bibnamefont {Lin}}, \bibinfo {author} {\bibfnamefont {P.}~\bibnamefont
  {Bourges}}, \bibinfo {author} {\bibfnamefont {B.}~\bibnamefont {Keimer}}, \
  and\ \bibinfo {author} {\bibfnamefont {Y.}~\bibnamefont {Sidis}},\ }\href
  {http://link.aps.org/doi/10.1103/PhysRevLett.105.037207} {\bibfield
  {journal} {\bibinfo  {journal} {Phys. Rev. Lett.},\ }\textbf {\bibinfo
  {volume} {105}},\ \bibinfo {pages} {037207} (\bibinfo {year}
  {2010})}\BibitemShut {NoStop}%
\end{thebibliography}
%

\end{document}